\documentclass[11pt,twoslides,fleqn]{scrartcl}

\usepackage{caption}
\usepackage{float}
\usepackage{subcaption}
\usepackage{amssymb}
\usepackage{graphicx}
\usepackage[utf8]{inputenc}
\usepackage[T1]{fontenc}
\usepackage{amsmath}
\usepackage{curves}
\usepackage{textcomp} 
\usepackage[normalem]{ulem}
\usepackage{rotating}
\usepackage[margin=2cm,outer=3cm]{geometry}
\usepackage{booktabs}
\usepackage{tabularx}
\usepackage{siunitx}
\usepackage[version=4]{mhchem}
\usepackage{listings}
\usepackage{authblk}

\usepackage{color}
\usepackage{nomencl}
\usepackage{bm}
\usepackage{graphicx}
\usepackage{subcaption}
\usepackage{tikz}
\usepackage{stanli}
\tikzstyle{load}   = [ultra thick,-latex]
\tikzstyle{stress} = [-latex]
\tikzstyle{dim}    = [latex-latex]
\tikzstyle{axis}   = [-latex,black!55]
\usetikzlibrary{shapes,arrows.meta,calc}

\usepackage{lineno}

\usepackage{times}

\usepackage[colorlinks=true,allcolors=blue]{hyperref}

\DeclareSIUnit\month{month}
\DeclareMathOperator{\Penalty}{Penalty}
\usepackage{csquotes}

\usepackage[authoryear]{natbib}

\usepackage{xcolor} 
\usepackage{amsmath}
\usepackage{tcolorbox} 
\graphicspath{{Foto/}}
\usepackage{epstopdf}

\usepackage{todonotes}

\begin{document}

\renewcommand\Affilfont{\fontsize{11}{11}\itshape}
\renewcommand\Authfont{\fontsize{11}{11}\itshape}
\title{A finite element-based machine learning model for hydro-mechanical analysis of swelling behavior in clay-sulfate rocks}

\author[1]{Reza Taherdangkoo}
\author[2]{Mostafa Mollaali}
\author[3]{Matthias Ehrhardt}
\author[1]{Thomas Nagel}
\author[4]{Lyesse Laloui}
\author[4]{Alessio Ferrari}
\author[1]{Christoph Butscher}

\affil[1]{TU Bergakademie Freiberg, Geotechnical Institute, Gustav-Zeuner-Str. 1, 09599 Freiberg, Germany}
\affil[2]{Helmholtz Centre for Environmental Research– UFZ, Department of Environmental Informatics, Leipzig, 04318, Germany}
\affil[3]{Applied and Computational Mathematics, Bergische Universität Wuppertal, Wuppertal, Germany}
\affil[4]{École Polytechnique Fédérale de Lausanne (EPFL), Laboratory of Soil Mechanics, Lausanne, Switzerland}

\date{}
\maketitle
{{\noindent \\Corresponding author, \url{Reza.Taherdangkoo@ifgt.tu-freiberg.de}}}

\section*{Highlights}

\begin{itemize}
   \item A hybrid constrained machine learning framework was developed for efficient prediction of swelling behavior in clay-sulfate rocks. 
   \item Penalty terms in the machine learning model ensured physically consistent predictions for porosity, saturation, and displacement. 
   \item Sensitivity analysis identified critical parameters, and Monte Carlo simulations confirmed robust predictions under uncertainty. 
\end{itemize}

\section*{Abstract}
The hydro-mechanical behavior of clay-sulfate rocks, especially their swelling properties, poses significant challenges in geotechnical engineering. 
This study presents a hybrid constrained machine learning (ML) model developed using the categorical boosting algorithm (CatBoost) tuned with a Bayesian optimization algorithm to predict and analyze the swelling behavior of these complex geological materials. Initially, a coupled hydro-mechanical model based on the Richards' equation coupled to a deformation process with linear kinematics implemented within the finite element framework OpenGeoSys was used to simulate the observed ground heave in Staufen, Germany, caused by water inflow into the clay-sulfate bearing Triassic Grabfeld Formation. A systematic parametric analysis using Gaussian distributions of key parameters, including Young's modulus, Poisson's ratio, maximum swelling pressure, permeability, and air entry pressure, was performed to construct a synthetic database. The ML model takes time, spatial coordinates, and these parameter values as inputs, while water saturation, porosity, and vertical displacement are outputs. In addition, penalty terms were incorporated into the CatBoost objective function to enforce physically meaningful predictions. Results show that the hybrid approach effectively captures the nonlinear and dynamic interactions that govern hydro-mechanical processes. The study demonstrates the ability of the model to predict the swelling behavior of clay-sulfate rocks, providing a robust tool for risk assessment and management in affected regions. The results highlight the potential of ML-driven models to address complex geotechnical challenges.

\hfill \break
\textbf{Keywords}: clay-sulfate rocks; swelling; hydro-mechanical modeling; physics-based machine learning; categorical boosting

\clearpage

\section{Introduction}

The swelling behavior of clay-sulfate rocks is a major challenge in geotechnical engineering. 
This phenomenon is most commonly observed in geological formations that contain both clay minerals and anhydrite, such as the Triassic Grabfeld Formation (formerly known as the "Gypsum Keuper") in southern Germany and Switzerland \citep{Butscher2016, taherdangkoo2023hydro}. 
The swelling is triggered by water infiltration, leading to two distinct but interrelated mechanisms: the interlayer hydration and osmotic swelling of clay minerals, and the transformation of anhydrite (\ce{CaSO4}) into gypsum (\ce{CaSO4.2H2O}) in a hydration reaction. 
The latter mechanism, known as gypsification, is particularly important because it is associated with an increase in volume of up to 61\,\%. 
This expansion induces significant swelling pressures that often exceed the bearing capacity of surrounding soils or structures, leading to ground heave, structural damage, and failure of the affected infrastructure \citep{Madsen1989, Butscher2016}.

The complexity of swelling in clay-sulfate rocks results from the coupled interactions between hydraulic, mechanical, and chemical processes. 
For example, the chemical transformation of anhydrite to gypsum changes the pore structure, affecting permeability and porosity. 
This change, in turn, affects the hydraulic response of the rock, influencing the dynamics of water flow. 
At the same time, the induced swelling pressures cause mechanical deformations that alter the hydraulic behavior \citep{Wanninger2020, schweizer2018reactive}. 
These coupled processes are nonlinear and time-dependent, often spanning years or decades \citep{Wittke2014, taherdangkoo2023hydro}. 
In addition, self-sealing behavior caused by gypsum coating on anhydrite inhibits mineral dissolution, introducing a feedback mechanism that makes swelling behavior difficult to predict. 
Environmental factors such as changes in groundwater flow, stress conditions, and pore water chemistry further exacerbate the uncertainty in predicting swelling response \citep{Jarzyna2022, butscher2011relation}. 

Despite decades of research, understanding and managing the swelling behavior of clay-sulfate rocks remains a complex task \citep{taherdangkoo2024comparative, alonso2023positive, Wanninger2020}. 
Numerical modeling, particularly finite element modeling (FEM), has been widely used to study the swelling of expansive geomaterials. 
FEM provides an established approach to solving the physics-based coupled hydro-mechanical-chemical equations that govern swelling, allowing the simulation of complex interactions with high fidelity. 
However, the computational cost of FEM simulations increases significantly when applied to large-scale systems or when the variability and uncertainty of input parameters must be considered \citep{Buchwald2020}. 
The need for repeated evaluations when performing sensitivity analyses, uncertainty quantification, and parameter optimization, often limits FEM-based simulations for many real-world applications \citep{Buchwald2024}.

The emergence of machine learning (ML) offers a promising alternative to address these limitations. 
ML models are particularly well suited for problems characterized by nonlinear relationships and high-dimensional parameter spaces, such as the swelling behavior of clay-sulfate rocks \citep{Burden2009,taherdangkoo2022modeling}. 
ML models can learn the underlying patterns and dependencies between input parameters and outputs by training on data sets generated from FEM simulations or field measurements \citep{Virupaksha2024}. 
Once trained, these models can provide predictions orders of magnitude faster than FEM, making them invaluable for real-time analysis, optimization, or risk assessment. 
However, a key challenge in applying ML to geotechnical problems is ensuring that the predictions remain physically consistent, as purely data-driven ML models run the risk of producing outputs that are physically implausible \citep{taherdangkoo2024, kooti2024constrained}.

This study addresses these challenges by developing a hybrid modeling framework that combines physics-based FEM with constrained ML to predict and analyze the swelling behavior of clay-sulfate rocks. 
The FEM model focuses on hydro-mechanical processes and incorporates swelling pressures as a function of water saturation to indirectly account for chemical contributions to volumetric strain, allowing efficient prediction of deformation and porosity changes. 
First, a coupled hydro-mechanical model -- based on Richards' equation and linear kinematics -- implemented within the OpenGeoSys framework was used to simulate the observed ground heave at Staufen, Germany. 
A systematic parametric analysis was performed to construct a synthetic database that was used to train a constrained machine learning model.
This model, developed using the categorical boosting algorithm (CatBoost) and optimized with Bayesian methods, integrates time, spatial coordinates, and material properties as inputs to predict the evolution of porosity, saturation, and vertical displacement during swelling.
We then quantified parametric uncertainties using Monte Carlo simulations.

\section{Problem statement}
The rock swelling occurred following the failure of geothermal drilling operations in 2007 in the town of Staufen in southwestern Germany. 
The wellbore failure resulted in upward flow of water from underlying artesian aquifers into the Grabfeld formation, a clay layer characterized by significant anhydrite content. 
The interaction of the water with sulfate minerals caused both mechanical swelling of the clay and chemical swelling through anhydrite gypsification. 
These processes led to a significant increase in volume of the rock mass, resulting in long-term ground deformations \citep{Grimm2014, Ruch2013, Schweizer2019}.

In response, a number of mitigation measures were implemented in 2009~\citep{LGRB2010}. These measures included the re-grouting of the defective boreholes and the installation of pumping wells in the artesian aquifer beneath the Grabfeld formation.
Their goal was to lower the hydraulic potential in the aquifer, thereby reducing water intrusion into the clay-sulfate rocks and mitigating further swelling. 
The Geological Survey of Baden-Württemberg (LGRB) established a geodetic network with 106 observation points. 
The ground surface displacement was recorded from January 2008 to September 2011, with each point being sampled up to 49 times at intervals of 11 to 63 days \citep{LGRB2010, LGRB2012}.

\section{Finite element method}
A machine learning model was developed based on a previously established hydro-mechanical (HM) finite element framework to simulate the swelling behavior of clay-sulfate rocks in the geological setting of Staufen, Germany \citep{taherdangkoo2022coupled}. 

\subsection{Mathematical model and numerical implementation}
A summary of the governing equations is presented here, with more comprehensive details available in \citet{taherdangkoo2022coupled}. The fluid flow within the porous media is governed by the Richards' equation~\citep{Kafle2022,Pitz2023}:
\begin{equation}
   \rho_\text{w}\left[\phi+\frac{p_\text{w}S_\text{w}(\alpha_\text{B}-\phi)}{K_\text{s}}\right]\dot{S}_\text{w}+\rho_\text{w}S_\text{w}\left[\frac{\phi}{K_\text{w}}+S_\text{w}\frac{\alpha_\text{B}-\phi}{K_\text{s}}\right]\dot{p}_\text{w}+\nabla \cdot \mathbf{q}+\alpha_\text{B}\rho_\text{w}S_\text{w}\nabla \cdot \mathbf{\dot{u}}=0,
\end{equation}
where $\phi$ is the porosity, $\rho_\text{w}$ is the water density, and $S_\text{w}$ is the water saturation, 
$p_\text{w}$ is the water pressure, $\mathbf{u}$ the displacement of the solid skeleton, and $\mathbf{q}$ is the Darcy mass flux. 

The Biot-Willis coefficient is $\alpha_\text{B}$, and the bulk moduli of the solid and water phase are $K_\text{s}$ and $K_\text{w}$, respectively. The relation between capillary pressure and saturation was described by \citet{vangenuchten}.

The mechanical equilibrium is described by:
\begin{equation}
    \nabla\cdot\boldsymbol\sigma+\left[\rho_\text{s}(1-\phi)+S_\text{w}\rho_\text{w}\phi \right]\mathbf{g}=\boldsymbol 0,
\end{equation}
where $\rho_\text{s}$ is the intrinsic solid density and $\boldsymbol\sigma$ is the total stress tensor. The relationship between the total stress tensor, $\boldsymbol\sigma$, and the effective stress tensor, $\boldsymbol\sigma'$, follows the extended Bishop's model:
\begin{equation}
    \boldsymbol\sigma = -\chi(S_\text{w}) p_\text{w} \mathbf{I} + \boldsymbol\sigma' - \sigma^\text{sw} \mathbf{I},
\end{equation}
where $\chi(S_\text{w})$ is Bishop's function, often set equal to the water saturation $S_\text{w}$.
The unit tensor is denoted by $\mathbf{I}$, and $\sigma^\text{sw}$ represents the swelling pressure. The swelling behavior is described by relating the swelling pressure $\sigma^\text{sw}$ to the maximum swelling pressure, $\sigma^\text{sw}_\text{max}$, and to the change in water saturation via \citep{Chaudhry2024, taherdangkoo2022coupled}:
\begin{equation}\label{eq:8}
    \sigma^\text{sw} = \sigma^\text{sw}_\text{max}(S_\text{w}-S_\text{wi})
\end{equation}

The described processes were implemented using OpenGeoSys (OGS) \citep{Kolditz2012, Bilke2019},
a finite element-based simulator for solving coupled processes in porous media.

\subsection{Numerical model setup}

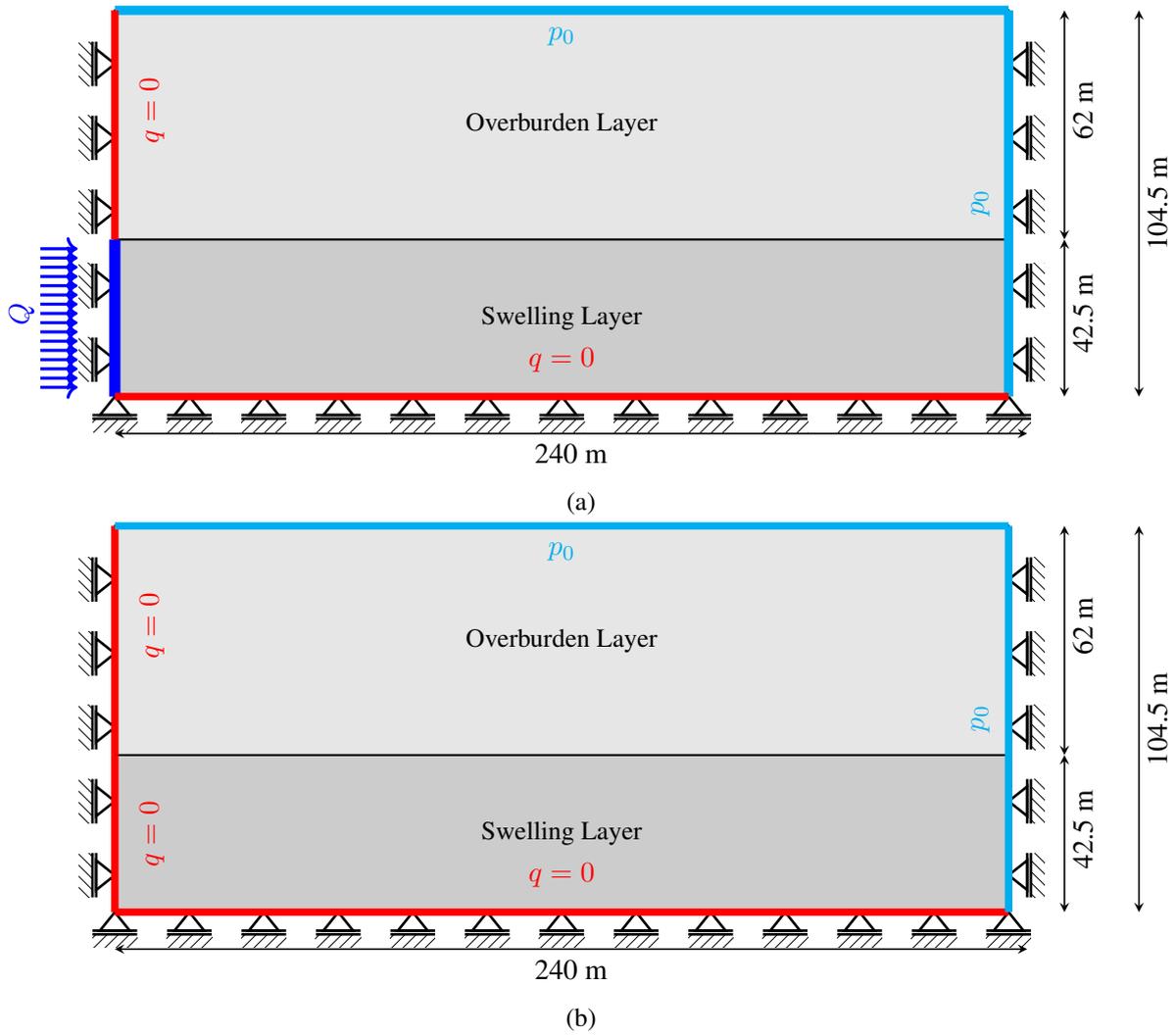
\begin{figure}[H]
    \begin{subfigure}[l]{1\textwidth}
        \raggedleft

        \begin{tikzpicture}[scale=0.5] 
            \fill[gray!20] (0, 4.25) rectangle (24, 10.45); 
            \fill[gray!40] (0, 0) rectangle (24, 4.25);     

            \draw[thick] (0, 0) rectangle (24, 10.45); 

            \draw[thick] (0, 4.25) -- (24, 4.25); 

            \node[align=center, font=\small] at (12, 2.125) {Swelling Layer};
            \node[align=center, font=\small] at (12, 7.35) {Overburden Layer};

            \draw[stealth-stealth, semithick] (25.5, 0) -- (25.5, 4.25) node[black, align=center, rotate=90] at (26, 2.25) {42.5 m};
            \draw[stealth-stealth, semithick] (25.5, 4.25) -- (25.5, 10.45) node[black, align=center, rotate=90] at (26, 7.75) {62 m};
            \draw[stealth-stealth, semithick] (27.5, 0) -- (27.5, 10.45) node[black, align=center, rotate=90] at (28, 10.45/2) {104.5 m};

            \draw[stealth-stealth, semithick] (0, -1) -- (24.5, -1) node[midway, below] {240 m};

            \foreach \i in {0,1,...,12} {
                \point{a\i}{\i*1}{0}; 
                \support{2}{a\i};      
            }

            \foreach \i in {1,...,5} {
                \point{l\i}{12}{\i*1-0.5}; 
                \support{2}{l\i}[90];   
            }

            \foreach \i in {1,...,5} {
                \point{r\i}{0}{\i*1-0.5}; 
                \support{2}{r\i}[-90];     
            }

            \foreach \y in {0.25,0.5,...,4.0}
                \draw[->, very thick, blue] (-2, \y) -- (-1., \y);
            \node[blue, rotate=90, align=center] at (-2.5, 2.25) {$Q$};

            \draw[line width=1.5mm,blue] (0,0) -- (0,4.25); 

            \draw[line width=1.25mm,cyan] (24,0) -- (24,10.45); 
            \node[cyan, rotate=0, align=center] at (12, 9.75) {$p_0$};

            \draw[line width=1.25mm,cyan] (0,10.45) -- (24,10.45); 
            \node[cyan, rotate=90, align=center] at (23.25, 5.25) {$p_0$};

            \draw[line width=1.mm,red] (0,0) -- (24,0); 
            \node[red, rotate=0, align=center] at (12, 1) {$q=0$};

            \draw[line width=1.mm,red] (0,4.25) -- (0, 10.45); 
            \node[red, rotate=90, align=center] at (1, 7.75) {$q=0$};
            
        \end{tikzpicture}
        \caption{}
        \label{fig:schematic1}
    \end{subfigure}
    \vfill
    \begin{subfigure}[l]{1\textwidth}
    \raggedleft
        \begin{tikzpicture}[scale=0.5] 
            \fill[gray!20] (0, 4.25) rectangle (24, 10.45); 
            \fill[gray!40] (0, 0) rectangle (24, 4.25);     

            \draw[thick] (0, 0) rectangle (24, 10.45); 

            \draw[thick] (0, 4.25) -- (24, 4.25); 

            \node[align=center, font=\small] at (12, 2.125) {Swelling Layer};
            \node[align=center, font=\small] at (12, 7.35) {Overburden Layer};

            \draw[stealth-stealth, semithick] (25.5, 0) -- (25.5, 4.25) node[black, align=center, rotate=90] at (26, 2.25) {42.5 m};
            \draw[stealth-stealth, semithick] (25.5, 4.25) -- (25.5, 10.45) node[black, align=center, rotate=90] at (26, 7.75) {62 m};
            \draw[stealth-stealth, semithick] (27.5, 0) -- (27.5, 10.45) node[black, align=center, rotate=90] at (28, 10.45/2) {104.5 m};

            \draw[stealth-stealth, semithick] (0, -1) -- (24.5, -1) node[midway, below] {240 m};

            \foreach \i in {0,1,...,12} {
                \point{a\i}{\i*1}{0}; 
                \support{2}{a\i};      
            }

            \foreach \i in {1,...,5} {
                \point{l\i}{12}{\i*1-0.5}; 
                \support{2}{l\i}[90];   
            }

            \foreach \i in {1,...,5} {
                \point{r\i}{0}{\i*1-0.5}; 
                \support{2}{r\i}[-90];   
            }

            \draw[line width=1.mm,red] (0,0) -- (0,4.25); 
            \node[red, rotate=90, align=center] at (1, 2.125) {$q=0$};

            \draw[line width=1.mm,red] (0,0) -- (24,0); 
            \node[red, rotate=0, align=center] at (12, 1) {$q=0$};

            \draw[line width=1.mm,red] (0,4.25) -- (0, 10.45); 
            \node[red, rotate=90, align=center] at (1, 7.75) {$q=0$};

            \draw[line width=1.mm,cyan] (24,0) -- (24,10.45); 
            \node[cyan, rotate=0, align=center] at (12, 9.75) {$p_0$};

            \draw[line width=1.mm,cyan] (0,10.45) -- (24,10.45); 
            \node[cyan, rotate=90, align=center] at (23.25, 5.25) {$p_0$};
        \end{tikzpicture}
        \caption{}
        \label{fig:schematic2}
    \end{subfigure}
    \caption{A schematic view of the numerical model illustrating the flow and mechanical boundary conditions during (a) the leakage period and (b) the mitigation period.}
    \label{fig:schematic}
\end{figure}

The axisymmetric 2D domain extends horizontally for \SI{240}{m}  and vertically for \SI{104.5}{m}, and includes two layers: the swelling layer (\SI{42.5}{m}) and the overburden layer (\SI{62}{m}) Figure~\ref{fig:schematic}.  
The overburden layer represents the consolidation of all layers above the swelling layer into a single equivalent layer. 
The sedimentary layers are treated as homogeneous throughout the domain.

The simulation proceeds in three phases: 
(1) establishment of a steady-state initial condition, 
(2) water inflow into the swelling layer following geothermal drilling on $3$ September $2007$, and 
(3) stopping of the water inflow as a result of mitigation measures initiated on $4$ November $2009$. 
The total simulation time for phases (2) and (3) was $1490$ days, with $790$ days for the leakage period and $710$ days for the mitigation phase~\citep{LGRB2010,LGRB2012}.

During the leakage phase, water entered the swelling layer (\SI{-104.5}{m} $\leq y \leq$ \SI{-62}{m}) through the left boundary at a constant inflow rate of \SI{1.3e-1}{\kilo\gram\per\second} \citep{schweizer2018reactive}. The inflow was distributed over the boundary using a normalized Neumann flux boundary condition, with the normalization based on the borehole radius to represent the effective cross-sectional area for water entry into the domain.
In the mitigation phase ($710$ days), the inflow was stopped and this boundary was switched to a no-flow condition. 
The right boundary (\SI{-104.5}{m} $\leq y \leq$ \SI{0}{m}) was kept at the initial pressure, while the top boundary was defined as a free drainage surface and kept constant at its initial pressure. 
The model assumes pure water with a density of \SI{1000}{\kilo\gram\per\cubic\meter} for both ambient and inflowing water, despite known variations in groundwater composition~\citep{LGRB2010}. 
Mechanically, the lateral and bottom boundaries were fixed in the normal direction, while the top boundary was traction-free.

The elastic and hydrogeological parameters were obtained from \citep{BENZ2012,LGRB2010,LGRB2012,schweizer2018reactive}. 
The properties of the overburden layer were calculated as a weighted arithmetic mean of the geological layers. 
The material parameters used in the simulations are listed in \autoref{tab:material}.

\begin{table}[ht]
    \centering
    \caption{Material parameter values used in the finite element model.}
    \begin{tabular}{l|l|r|r}
        \hline    
        Property & Units & Swelling layer & Overburden layer \\\hline
        Young's modulus $(E)$ & \si{\mega\pascal} & {\num{500}} & {\num{1000}} \\
        Poisson's ratio $(\nu)$ & - & {\num{0.2}} & {\num{0.2}} \\
        Biot-Willis coefficient $(\alpha_\text{B})$ & - &  {\num{1}} & {\num{1}} \\
        Maximum swelling pressure ($\sigma_\text{max}^\text{sw})$ & \si{\mega\pascal} & {\num{8}} & {\num{0}}\\
        Porosity $(\phi)$ & - & {\num{0.077}} & {\num{0.14}}\\
        Intrinsic permeability $(k)$ & $\si{\meter\squared}$ &  {\num{2e-13}} &  {\num{8e-13}}\\
        Solid density $(\rho_\text{s})$ & \si{\kg\per\cubic\meter} & {\num{2670}} & {\num{2627}} \\
        Water Density $(\rho_\text{w})$ & \si{\kg\per\cubic\meter} & {\num{1000}} & {\num{1000}} \\
        Water viscosity $(\mu_\text{w})$ & \si{\pascal \cdot \second} & {\num{1e-3}} & {\num{1e-3}} \\
        Initial water saturation ($S_\text{wi}$) & - & {\num{0.13}} & {\num{0.13}} \\
        Van Genuchten parameter $(\text{m})$ & - & {\num{0.75}} & {\num{0.75}} \\
        Van Genuchten air entry pressure $(\text{p}_\text{b})$ & \si{\pascal} & {\num{2000}} & {\num{2000}}\\\hline
    \end{tabular}
    \label{tab:material}
\end{table}

\subsection{Sensitivity analysis}
A one-variable-at-a-time (OVAT) sensitivity analysis was performed to generate a synthetic data set for machine learning modeling. 
The analysis involved systematically varying key parameters of the clay-sulfate layer, including Young's modulus ($E$), Poisson's ratio ($\nu$), maximum swelling pressure ($\sigma_{sw}^{\text{max}}$), permeability ($k$), and air entry pressure ($p_\text{b}$). 
Throughout the analysis, the material properties of the overburden were held constant. The base values and corresponding ranges for these parameters are shown in Table~\ref{tab:sensitivity_analysis}.

\begin{table}[ht]
    \centering
    \caption{Parameter values used for the sensitivity analysis of the swelling layer.}
    \begin{tabular}{l|l|r|r|r}
        \hline    
        Parameters & Units & Base Value & Minimum & Maximum \\\hline
        Young's modulus ($E$) & $\si{\mega\pascal}$ & \num{500} & \num{300} & \num{2500} \\
        Poisson's ratio ($\nu$) & - & \num{0.2} & \num{0.16} & \num{0.35} \\
        Swelling pressure ($\sigma_\text{max}^\text{sw}$) & $\si{\mega\pascal}$ & \num{8} & \num{3.2} & \num{13} \\
        Permeability ($k$) & $\si{\meter\squared}$ & \num{2e-13} & \num{1e-14} & \num{1e-12} \\
        Air entry pressure ($p_\text{b}$) & $\si{\pascal}$ & \num{2000} & \num{1000} & \num{3500} \\\hline
    \end{tabular}
    \label{tab:sensitivity_analysis}
\end{table}

To systematically explore the parameter space, $30$ values were generated for each parameter using Gaussian distributions centered on their base values.
The standard deviations for these distributions were determined as follows:
\begin{equation}
      s_{\text{param}} = \frac{\text{max}_{\text{param}} - \text{min}_{\text{param}}}{3},
\end{equation}
where $\text{max}_{\text{param}}$ and $\text{min}_{\text{param}}$ represent the maximum and minimum values of each parameter's range, respectively. 
This approach resulted in a total of $150$ parameter sets, which were then used to perform simulation runs.

\section{Machine learning model}

\subsection{CatBoost algorithm}
The machine learning algorithm used, CatBoost, is a gradient boosting method designed for efficient handling of categorical features and improved generalization. 
It iteratively constructs an ensemble of decision trees, where each tree is trained to minimize the residual errors of the preceding trees. 
For regression tasks, the general objective function in gradient boosting can be written as~\citep{dorogush2018catboost, prokhorenkova2018catboost}:
\begin{equation}
     \mathcal{L}(\mathbf{y}, \mathbf{\hat{y}}) = \sum_{i=1}^n \ell(y_i, \hat{y}_i) + \Omega(\Theta),
\end{equation}
where $y_i$ is the true value for the $i^\text{th}$ sample obtained from the finite element analysis, $\hat{y}_i$ is the predicted value, and $\ell(y_i, \hat{y}_i)$ denotes the loss function. 
The regularization term $\Omega(\Theta)$ depends on the model parameters $\Theta$ and is used to prevent overfitting. At each iteration $t$, the model fits a decision tree $h_t(x)$ to the negative gradient of the loss function:
\begin{equation} h_t(x) \approx g_t = -\nabla_{\hat{y}_{t-1}} \mathcal{L}(y, \hat{y}_{t-1}),
\end{equation}
where $\mathbf{g}_t$ is the decent direction for the boosting step $t$. 
The predictions are then updated as: \begin{equation} 
     \hat{y}_t = \hat{y}_{t-1} + \eta \cdot h_t(x),
\end{equation} 
where $\eta$ is the learning rate, controlling the contribution of each tree. 
To enforce physically meaningful predictions, the standard loss function (e.g., mean squared error, MSE) is modified with custom penalties.
The penalized loss function is defined as
\begin{equation}
    \mathcal{L}_{\text{total}} = \frac{1}{n} \sum_{i=1}^n \left[ (y_i - \hat{y}_i)^2 + \alpha \cdot \Penalty(y_i, \hat{y}_i) \right],
\end{equation}
where $\alpha$ is the penalty scale. Linear, exponential, or quadratic penalties were applied separately to each output variable -- porosity, saturation, and displacement -- with the threshold $\tau > 0$ adjusted based on the magnitude of the residuals for each variable. 
The penalty functions are defined as follows
\begin{equation}
     \Penalty_{\text{linear}} = \frac{1}{n} \sum_{i=1}^n \max\big(0, \alpha (|y_i - \hat{y}_i| - \tau)\big),
\end{equation}
\begin{equation}
     \Penalty_{\text{quadratic}} = \frac{1}{n} \sum_{i=1}^n \max\big(0, \alpha (|y_i - \hat{y}_i| - \tau)^2\big),
\end{equation}
\begin{equation}
    \Penalty_{\text{exponential}} = \frac{1}{n} \sum_{i=1}^n \max\big(0, e^{\alpha (|y_i - \hat{y}_i| - \tau)} - 1\big),
\end{equation}

Bayesian optimization was used to tune the hyperparameters, including the number of trees (n\_estimators), the maximum tree depth (max\_depth), the learning rate ($\eta$), the L2 regularization term ($\lambda$), and the penalty scale ($\alpha$). 
Optimal configurations were determined separately for each output variable to effectively train the CatBoost models.

\subsection{Bayesian optimization}
Bayesian optimization was used to systematically tune the hyperparameters and minimize the penalized loss function $\mathcal{L}_{\text{total}}$.
The surrogate model, $\mathcal{P}(\mathcal{O}(\theta))$, was modeled as a Gaussian process (GP) \citep{foresee1997gauss, snoek2012practical,taherdangkoo2021gaussian}:
\begin{equation}
    \mathcal{P}(\mathcal{O}(\theta)) \sim \mathcal{GP}(\mu(\theta), k(\theta, \theta')),
\end{equation}
where $\mu(\theta)$ is the mean function and $k(\theta, \theta')$ is the kernel function that captures the similarity between two hyperparameter configurations $\theta$ and $\theta'$. 
To select the next set of hyperparameters to evaluate, the expected improvement ($EI$) detection function was used, written as
\begin{equation}
     \text{EI}(\theta) = \mathbb{E}\left[ \max(0, \mathcal{O}(\theta) - \mathcal{O}_{\text{best}}) \right],
\end{equation}
where $\mathcal{O}_{\text{best}}$ is the best objective value observed so far. The EI balances the trade-off between exploring regions of high uncertainty and exploiting regions that are likely to yield improvements near the current optimum.

The optimization process was initialized with 10 randomly sampled configurations to construct an initial GP surrogate model. 
The model was then iteratively refined over 140 optimization steps to converge on the best set of hyperparameters. 
During optimization, penalty contributions were tracked, allowing analysis of the impact of each output variable on the penalized loss function. 
The objective function minimized in this process was the negative penalized loss
\begin{equation}
     \mathcal{O}(\theta) = -\mathcal{L}_{\text{total}}(\mathbf{y}, \mathbf{\hat{y}}).
\end{equation}

\section{Monte Carlo simulation}
Monte Carlo simulation was used to quantify the uncertainty in the model predictions \citep{metropolis1949monte} using the trained ML model. 
A total of $N = \num{30000}$ samples were generated for $M = 5$ parameters: 
Young's modulus ($E$), Poisson's ratio ($\nu$), permeability ($K$), air entry pressure ($p_b$), and maximum swelling pressure ($\sigma_\text{max}^\text{sw}$). 
The scaled value for the $i^\text{th}$ sample and the $j^\text{th}$ parameter was calculated as follows:
\begin{equation}
     x_{\text{scaled}, i, j} = x_{\min, j} + (x_{\max, j} - x_{\min, j}) \cdot x_{\text{random}, i, j},
\end{equation}
where $x_{\text{random}, i, j}$ is a random number uniformly sampled from $[0, 1]$, and $x_{\text{min}, j}$ and $x_{\text{max}, j}$ are the minimum and maximum values for the $j^\text{th}$ parameter, respectively. 
This scaling ensures a consistent distribution of the parameters across their respective ranges. 
In addition to these primary parameters, three auxiliary features (i.e.\ time and coordinate systems) were sampled directly from the dataset. Random sampling with replacement \citep{taherdangkoo2024} was performed to draw $N$ samples of the features $\text{time\_value},\, x_s,\, \text{and}\, y_s$
from the dataset according to
\begin{equation}
   \text{time\_value}_i, \quad x_{s,i}, \quad y_{s,i}, \quad i = 1, 2, \ldots, N.
\end{equation}

The final input matrix, $\mathbf{X}_{\text{final}}$, was constructed by combining the scaled Monte Carlo data with the randomly sampled auxiliary features:
\begin{equation*}
\mathbf{X}_{\text{final}} =
\begin{bmatrix}
\text{time\_value}_1 & E_1 & \nu_1 & k_1 & p_{b,1} & {\sigma_\text{max}^\text{sw}}_1 & x_{s,1} & y_{s,1} \\
\text{time\_value}_2 & E_2 & \nu_2 & k_2 & p_{b,2} & {\sigma_\text{max}^\text{sw}}_2 & x_{s,2} & y_{s,2} \\
\vdots               & \vdots & \vdots & \vdots & \vdots   & \vdots                  & \vdots   & \vdots   \\
\text{time\_value}_N & E_N & \nu_N & k_N & p_{b,N} & {\sigma_\text{max}^\text{sw}}_N & x_{s,N} & y_{s,N}
\end{bmatrix}.
\end{equation*}
Each row of $\mathbf{X}_{\text{final}}$ represents a complete feature vector for one Monte Carlo sample. 
This dataset was then used to analyze the variability of the model outputs, enabling quantification of the uncertainty associated with the predictions.

\section{Results and discussion}

\subsection{Model performance assessment}
\begin{figure}
\centering
\includegraphics[width=0.9\textwidth]{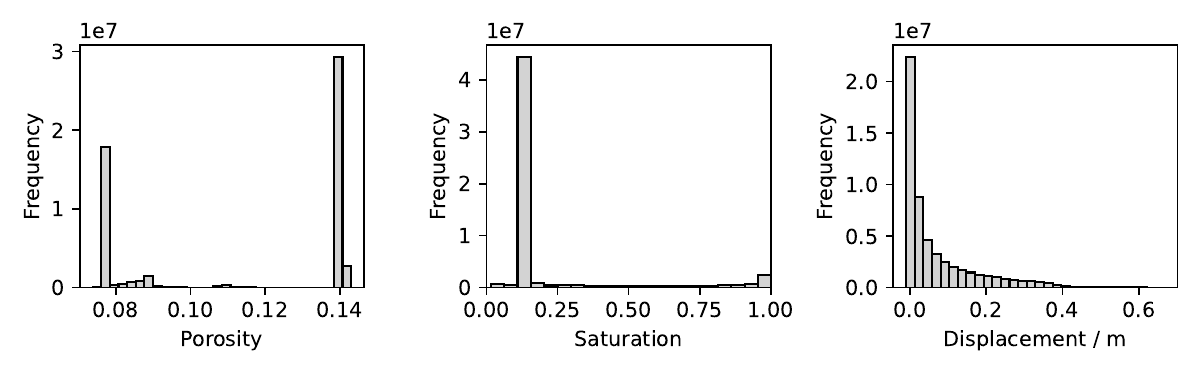}
\caption{Histograms showing the distribution of porosity, saturation, and displacement values for the entire dataset.}
\label{fig:distribution}
\end{figure}

The data set obtained from the sensitivity analysis contains approximately $5.4745 \cdot 10^{7}$ entries. It includes the parameters listed in Table~\ref{tab:sensitivity_analysis} along with porosity, saturation, and displacement values at 19209 finite element node and 25 time step for 114 simulations. The distributions of the aforementioned parameter values across the entire dataset are depicted in Figure~\ref{fig:distribution}.

Given the large size of the dataset, a $4:1$ splitting method was used to divide the data into training and test subsets.
This approach ensures computational efficiency while maintaining a sufficiently large and representative test set for model evaluation.
The random seed was set to $42$ for reproducibility, allowing consistent validation of results across runs. 

We ran several simulations to determine the optimal hyperparameter bounds of the CatBoost model, the number of iterations for the optimization process, the type of penalty function, and the size of the threshold $\tau$ for each output variable. 
The loss function was eventually modified with a linear penalty for porosity, an exponential penalty for saturation, and a quadratic penalty for displacement. 
A residual threshold of $0.001$ was applied uniformly to all three variables. 

\begin{figure}[H]
\centering
\includegraphics[width=0.7\textwidth]{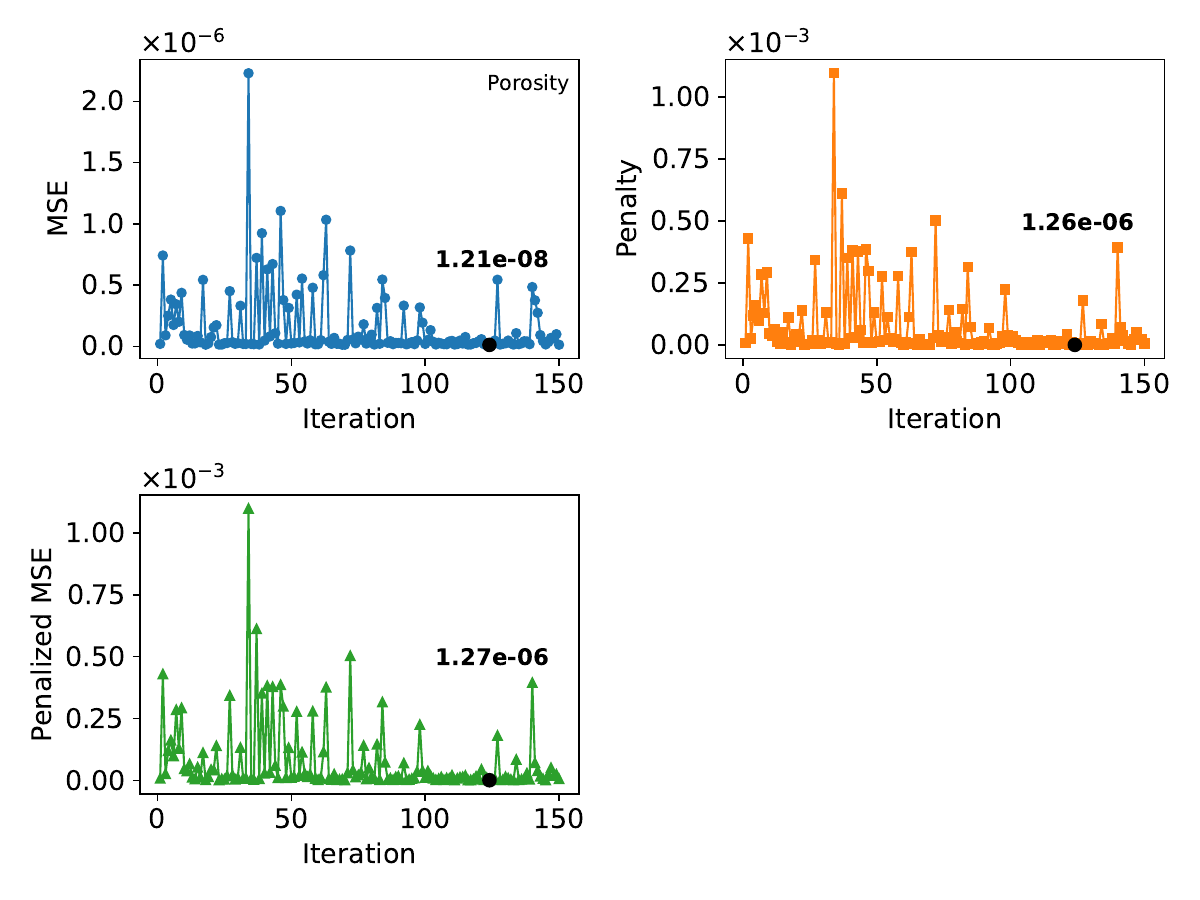}
\caption{The evolution of MSE, penalty term, and penalized MSE during hyperparameter tuning of the porosity model. 
Iteration 124 corresponds to the optimized parameter set.}
\label{fig:porosity_metrics}
\end{figure}

\begin{figure}[H]
\centering
\includegraphics[width=0.7\textwidth]{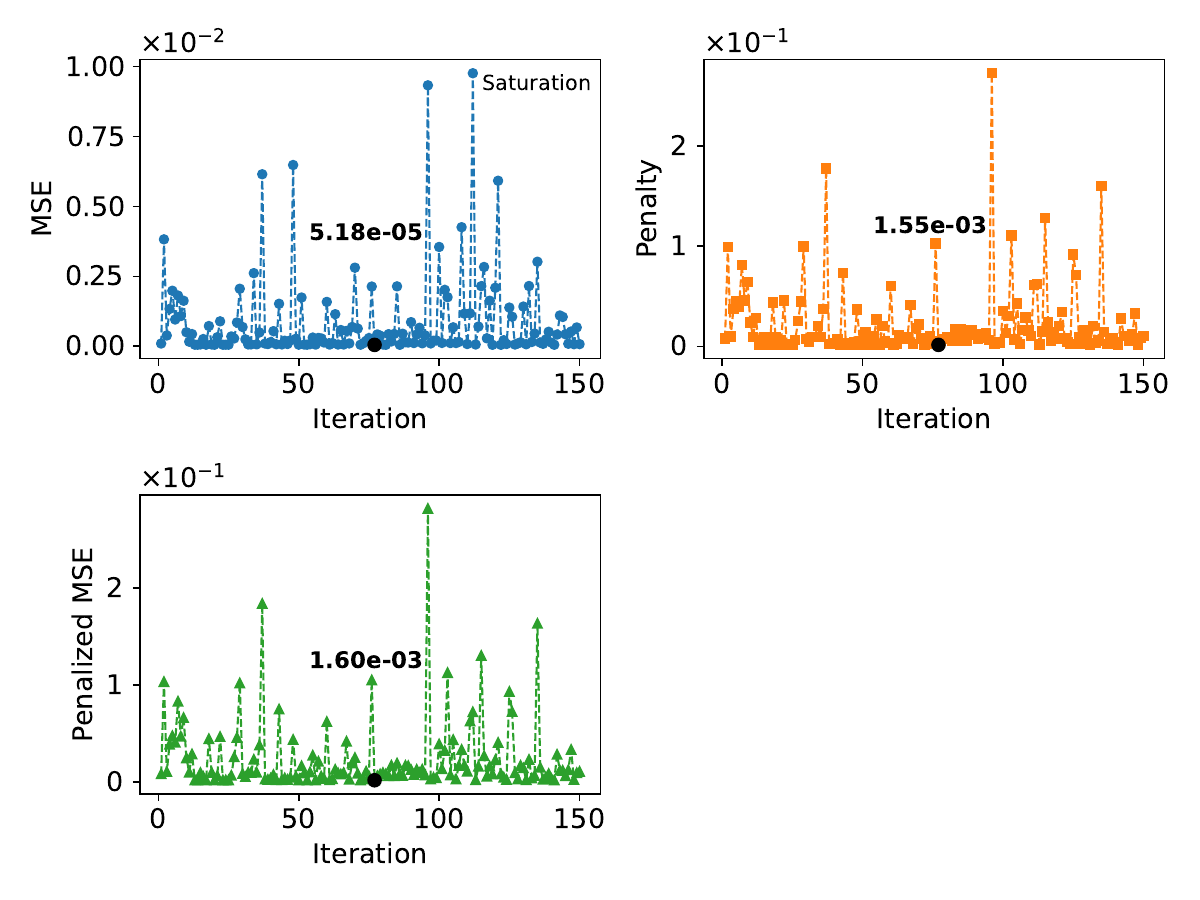}
\caption{The evolution of MSE, penalty term, and penalized MSE during hyperparameter tuning of the saturation model. 
Iteration 77 corresponds to the optimized parameter set.}
\label{fig:saturation_metrics}
\end{figure}

\begin{figure}[H]
\centering
\includegraphics[width=0.7\textwidth]{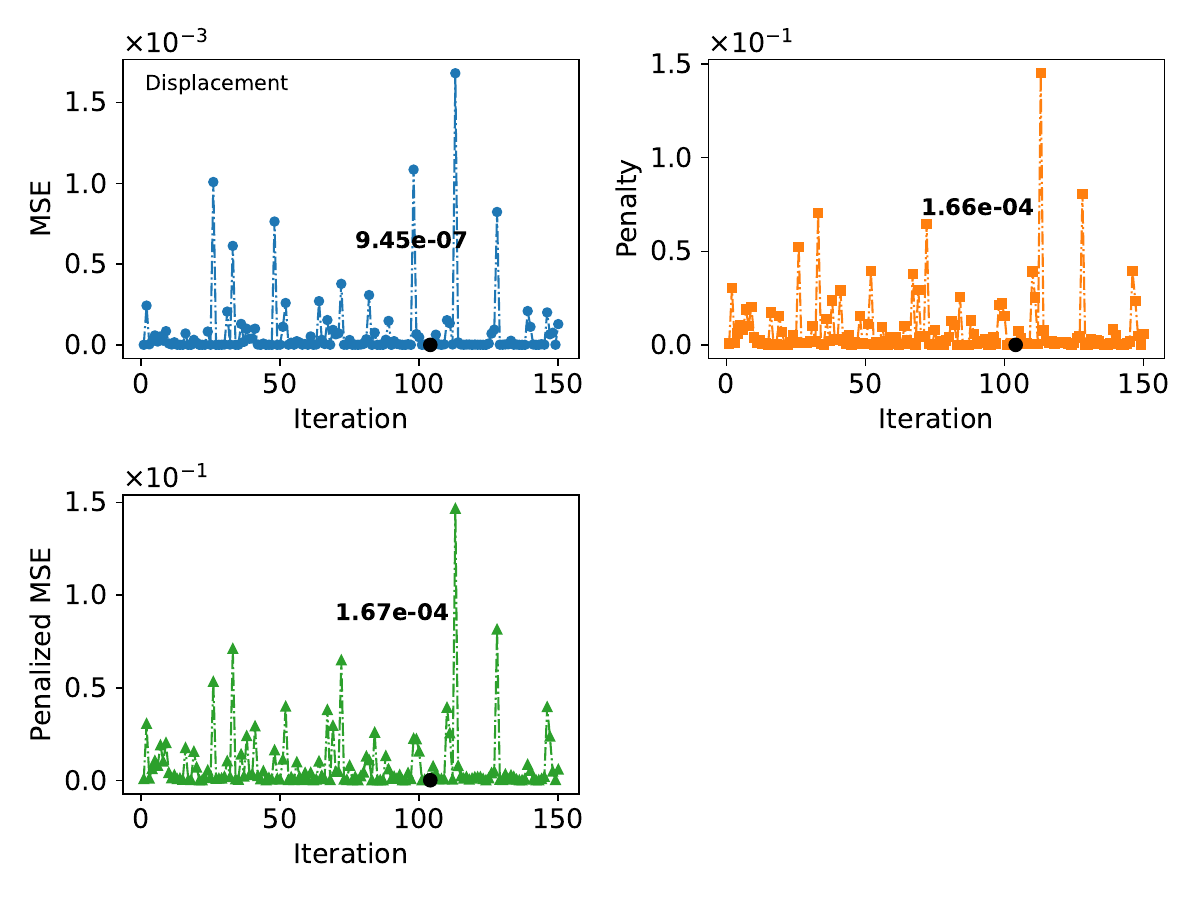}
\caption{The evolution of MSE, penalty term, and penalized MSE during hyperparameter tuning of the displacement model.
Iteration 104 corresponds to the optimized parameter set.}
\label{fig:displacement_metrics}
\end{figure}

Figures~\ref{fig:porosity_metrics}--\ref{fig:displacement_metrics} illustrate the Bayesian optimization of the porosity, saturation, and displacement models over $150$ iterations, and Table~\ref{hyperparams} lists the optimal hyperparameter configurations.
Figure~\ref{fig:porosity_metrics} shows high variability in early iterations, indicating exploration of the hyperparameter space. 
The porosity model stabilizes around iteration $120$, indicating convergence to an optimal solution.
Iteration $124$ achieves the lowest penalized MSE ($\num{1.272e-6}$), balancing model accuracy (MSE of $\num{1.208e-8}$) and regularization (penalty term of $\num{1.259e-6}$). 
The convergence pattern highlights the effectiveness of the optimization strategy in improving generalization.

The saturation model (Figure~\ref{fig:saturation_metrics}) optimizes at iteration $77$, with an MSE of $\num{5.177e-5}$, a penalty term of $\num{1.545e-3}$, and a penalized MSE of $\num{1.597e-3}$. 
This iteration demonstrates an optimal trade-off between prediction accuracy and regularization, as reflected in the minimized penalized MSE. 
The displacement model (Figure~\ref{fig:displacement_metrics}) optimizes at iteration $104$, yielding an MSE of $\num{9.458e-7}$, a penalty term of $\num{1.664e-4}$, and a penalized MSE of $\num{1.673 e-4}$. 
This iteration effectively balances accuracy and regularization, with minimal trade-off in predictive performance due to regularization.

Table~\ref{hyperparams} shows consistency in hyperparameters such as maximum depth ($16$) and learning rate ($0.25$), indicating that deeper trees and relatively higher learning rates are needed to capture the complex relationships in the data.
The regularization parameters $\lambda$ and $\alpha$ are fixed at $1.0$ for the saturation and displacement models,
indicating that minimal regularization is sufficient to achieve optimal performance in these models.
In contrast, the porosity model requires a higher $\lambda$ value ($6.0$), suggesting stronger regularization to avoid overfitting.
In addition, the number of estimators is nearly maximized for the saturation and displacement models ($594$ and $593$), while the porosity model stabilizes at a slightly lower value ($497$).

\begin{table}[h]
\caption{Hyperparameter ranges and optimized values for CatBoost models.}\label{hyperparams}
\centering
\begin{tabular}{@{}lcccc@{}}
\toprule
Hyperparameter         & Range           & \multicolumn{3}{c}{Models}                           \\ \cmidrule(lr){3-5}
                       &                 & Porosity         & Saturation          & Displacement      \\ \midrule
n\_estimators        & $100$ to $600$  & $497$              & $594$             & $593$             \\
max\_depth           & $3$ to $16$     & $16$               & $16$              & $16$              \\
$\eta$ & $0.01$ to $0.25$ & $0.25$             & $0.25$            & $0.25$            \\
$\lambda$ & $1$ to $6$      & $6.0$              & $1.0$             & $1.0$             \\
$\alpha$ & $1$ to $5$   & $1.0$              & $1.0$             & $1.0$             \\ \bottomrule
\end{tabular}
\end{table}

The model performance metrics in Table~\ref{model_metrics} show high prediction accuracy and generalization for all three models.
The porosity model achieved near-perfect $R^2$ values of $0.9999$ and a root mean square error (RMSE) of $0.0001$ for both training and test datasets. 
Similarly, the saturation model achieved $R^2$ values of $0.9991$ with RMSE values of $0.007$ (training) and $0.0072$ (test), indicating reliable performance even with potentially more variable data. 
The displacement model also achieved $R^2$ values of $0.9999$ and an RMSE of $\SI{0.001}{\meter}$ on both data sets, confirming its accuracy and compliance with residual constraints.

\begin{table}[h]
\caption{Model performance metrics.}\label{model_metrics}
\centering
\begin{tabular}{@{}lccccc@{}}
\toprule
Model        & \multicolumn{2}{c}{Train}      & \multicolumn{2}{c}{Test}       \\ \cmidrule(lr){2-3} \cmidrule(lr){4-5}
             & \( R^2 \)    & RMSE           & \( R^2 \)    & RMSE            \\ \midrule
Porosity (-)  & $0.9999$     & $0.0001$       & $0.9999$     & $0.0001$        \\
Saturation  (-)   & $0.9991$     & $0.007$       & $0.9991$     & $0.0072$        \\
displacement (\si{\meter}) & $0.9999$     & $0.001$       & $0.9999$     & $0.001$        \\ \bottomrule
\end{tabular}
\end{table}

The residuals of the porosity model for both training and test data sets exhibit a narrow range, 
with means close to zero and standard deviations of $0.0001$, demonstrating the model's ability to make accurate predictions with negligible bias. 
The saturation model shows a slightly wider residual variability, 
with training residuals ranging from $-0.665$ to $0.708$ and test residuals ranging from $-0.91$ to $0.969$. 
Despite the wider range, mean residuals of zero and standard deviations of $0.007$ indicate strong performance in capturing the complex relationships associated with saturation in both data sets. 

The displacement model has narrow residual ranges, confirming the model's effectiveness in handling quadratic penalties. 
The residuals are distributed around zero with standard deviations of $0.001$. 
Overall, the residual distributions validate the metrics in Table~\ref{model_metrics}, confirming the robustness of the framework. 
The consistency of near-zero residual means across models and datasets suggests minimal systematic bias, 
while the tight distributions demonstrate the ability of the models to generalize effectively.

\begin{figure}[H]
\centering
\includegraphics[width=0.75\textwidth]{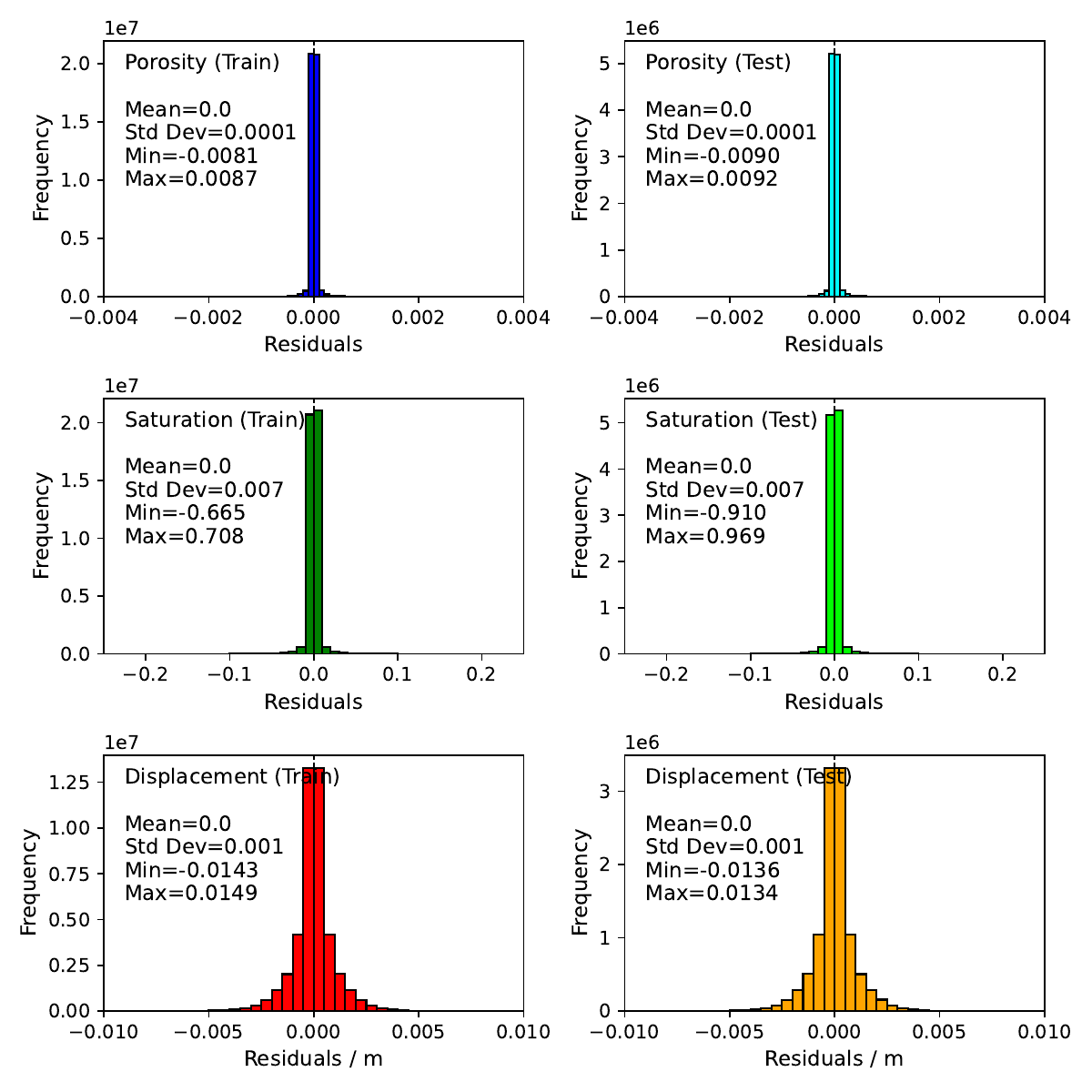}
\caption{Residual distributions for the train and test datasets of the saturation, porosity, and displacement models. }
\label{fig:residual}
\end{figure}

\begin{figure}[h]
\centering
\includegraphics[width=1\textwidth]{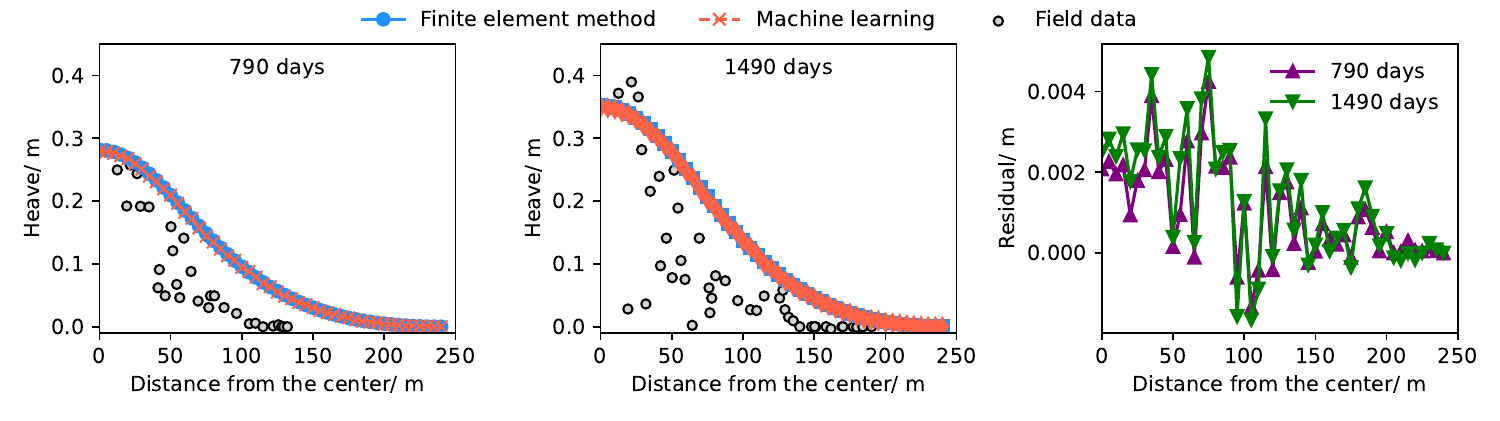}
\caption{Comparison of geodetic heave measurements with simulated heave using the finite element and machine learning models for two points in time: post-leakage ($790$ days) and post-mitigation ($1490$ days). The modeling residuals are shown.}
\label{fig:top}
\end{figure}

Figure~\ref{fig:top} shows the development of ground heave at post leakage ($790$~days) and post mitigation period ($1490$~days). 
The finite element and ML models replicate the shape and extent of ground surface heave, aligning with field observations from the geodetic monitoring network established by \citet{LGRB2010, LGRB2012}. 
The models are sufficiently accurate from a practical perspective. 
Discrepancies between the modeled and observed field data can be attributed to simplifying assumptions inherent in the finite element model.
For example, previous studies have shown that the gypsification of anhydrite, the primary process governing the swelling of clay-sulfate rocks, cannot be fully captured by hydro-mechanical models \citep{Wittke2014, taherdangkoo2023hydro}. 
In such models, swelling is driven by changes in the pore water saturation and once the pore space becomes saturated, the swelling process is considered completed. 
In reality, however, the complete transformation of anhydrite into gypsum rarely occurs. 

The ML model shows predictive accuracy comparable to the finite element method while offering significant computational efficiency.
The residuals, shown in the right panel of Figure~\ref{fig:top}, are primarily within $\pm \SI{0.003}{\metre}$ and show no discernible spatial bias. 
The residuals have near-zero means ($\SI{0.001}{\metre}$ at $790$~days and $\SI{0.0013}{\metre}$ at $1490$~days) as well as small standard deviations ($\SI{0.0012}{\metre}$ and $\SI{0.0015}{\metre}$).

\begin{figure}[h]
\centering
\includegraphics[width=1\textwidth]{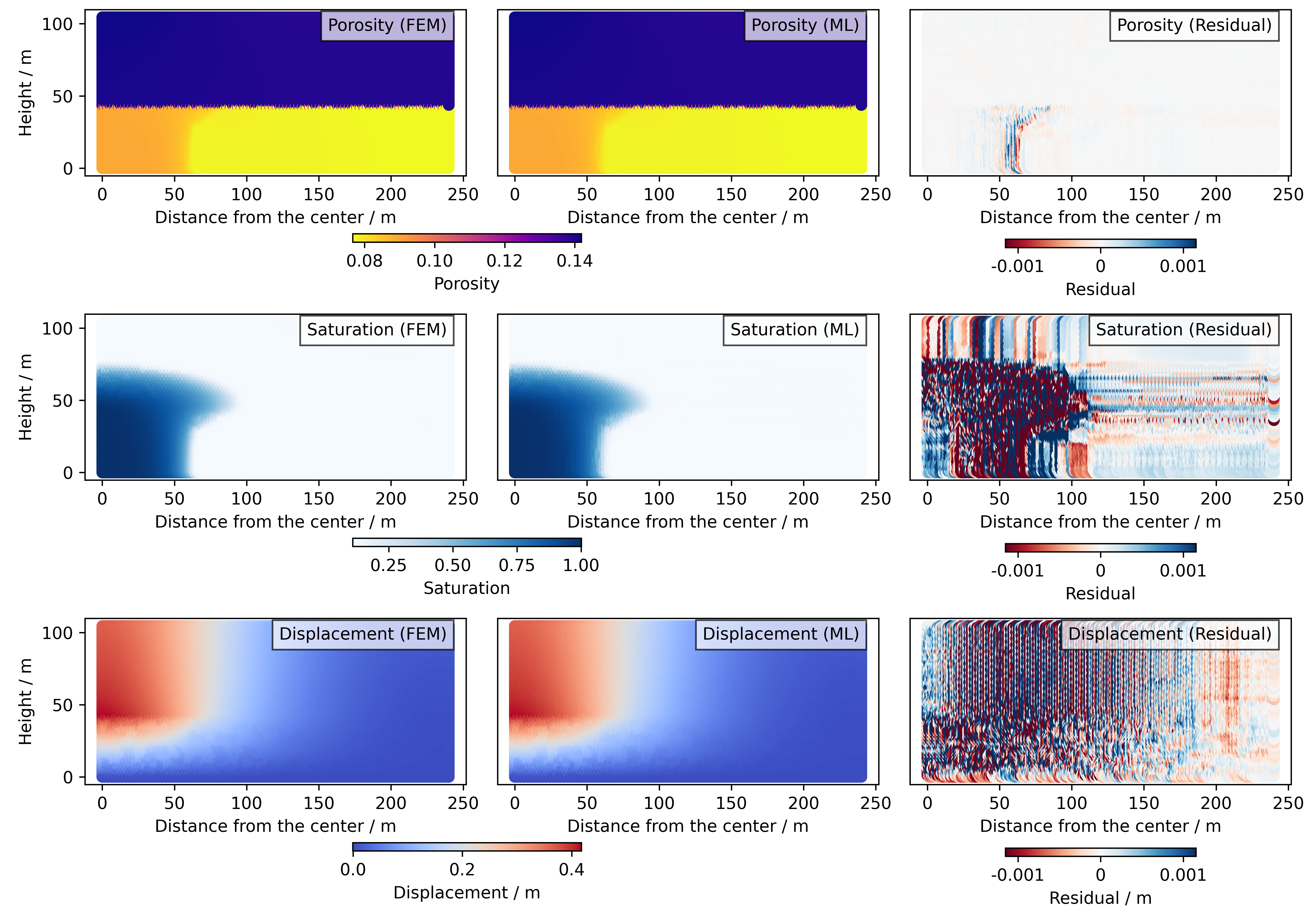}
\caption{Comparison of porosity, saturation, and displacement predicted by the finite element and the machine learning models across the spatial domain at $790$~day. 
The modeling residuals are depicted in the right panel.}
\label{fig:spatial_790}
\end{figure}

\begin{figure}[h]
\centering
\includegraphics[width=1\textwidth]{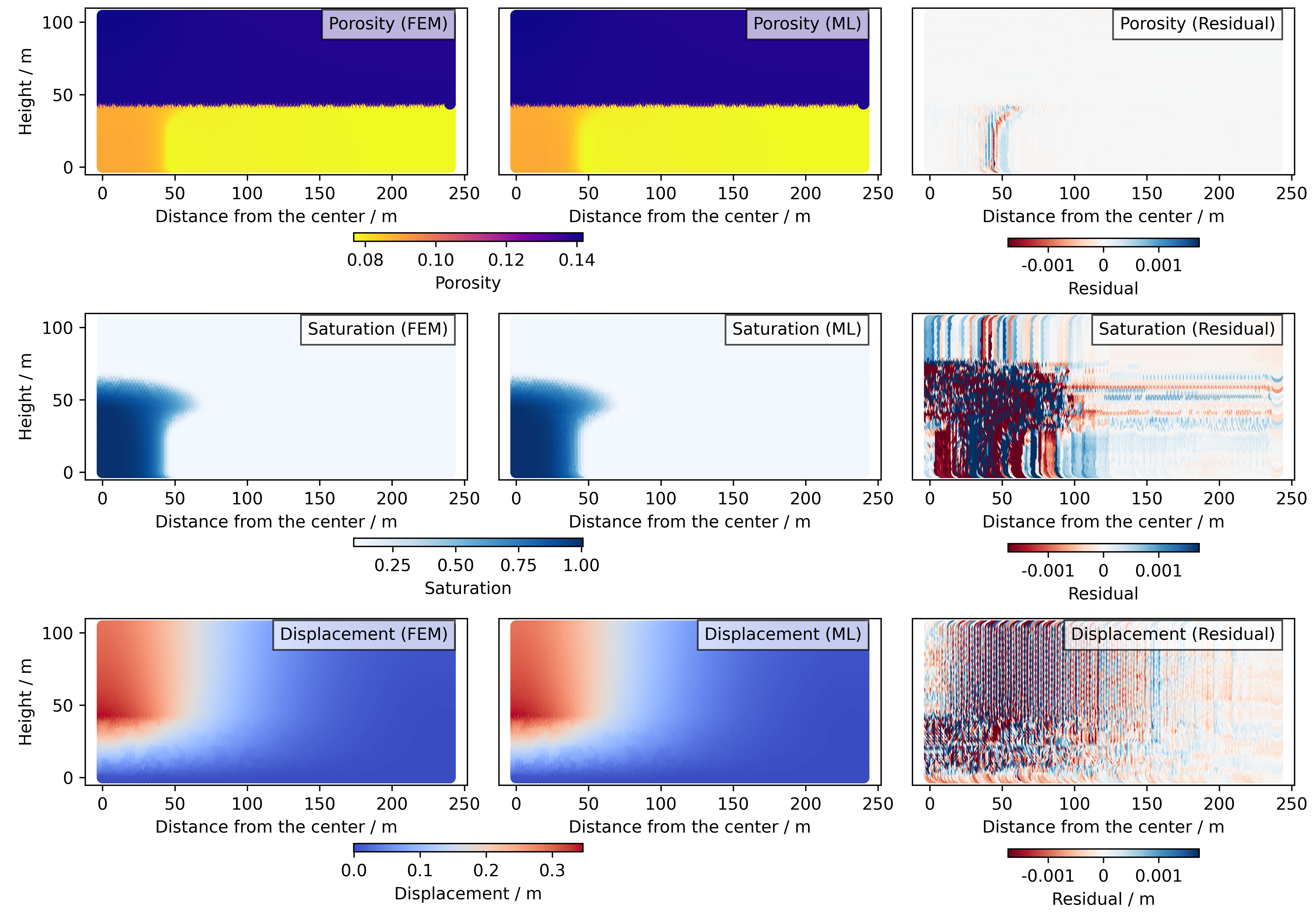}
\caption{Comparison of porosity, saturation, and displacement predicted by the finite element and the machine learning models across the spatial domain at $1490$~day.
The modeling residuals are depicted in the right panel.}
\label{fig:spatial_1490}
\end{figure}

Figure~\ref{fig:spatial_790} and Figure~\ref{fig:spatial_1490} show the spatial distributions of porosity, saturation, and displacement across the model domain at $790$ and $1490$ days.
Porosity predictions show minimal changes, consistent with the slow evolution of volumetric strain in clay-sulfate rocks and the uniform material properties of the swelling layer. 
Saturation distributions show the evolution of the water front due to water ingress, the primary trigger of swelling processes.
Displacement predictions accurately represent the swelling-induced ground heave, showing steep gradients in the vicinity of the swelling center that attenuate with distance due to the absence of swelling pressures in the far field. 
The comparison between the two time steps demonstrates the effectiveness of mitigation strategies as evidenced by the reduced water intrusion and associated swelling.

The porosity residuals show minimal spatial variation across the domain with the exception of the saturation front. 
The saturation residuals show greater variability in regions where water flow dynamics are more complex, such as near saturation fronts.
The displacement residuals are slightly larger near regions with steep displacement gradients. 
The analysis shows the limitations of the model in fully reproducing the FEM predictions under nonlinear conditions.
However, the residuals remain small, indicating that the model has high fidelity in replicating the nonlinear flow and mechanical responses.

\subsection{Sensitivity analysis}

\begin{figure}
\centering
\includegraphics[width=1\textwidth]{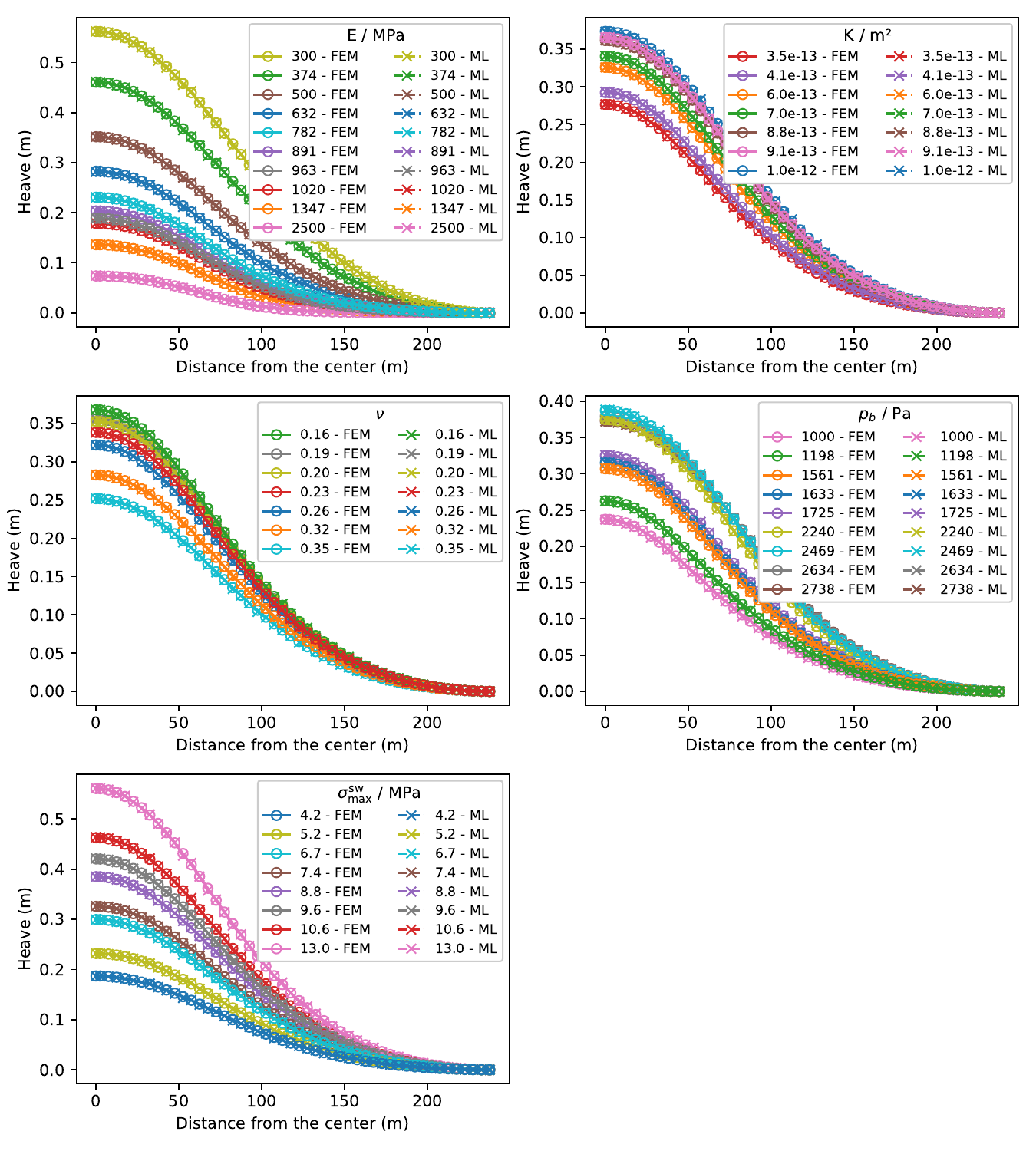}
\caption{Sensitivity analysis of various parameters on heave calculated from finite element and machine learning approaches. 
The heave at the ground surface at $1490$ day is plotted, varying Young’s modulus, permeability, Poisson’s ratio, air entry pressure, and maximum swelling pressure of the swelling layer.}
\label{fig:SA}
\end{figure}

Consistent with previous HM studies \citep{taherdangkoo2022coupled, schadlich2013application}, sensitivity analysis (Figure~\ref{fig:SA}) revealed that the model is most sensitive to changes in maximum swelling pressure and Young's modulus. 
The strong linear relationship between swelling pressure and displacement indicates the importance of accurately characterizing these parameters for predictive modeling. 
Poisson's ratio, permeability, and air entry pressure also influenced the model, although to a lesser extent. 
These results are consistent with previous parametric studies conducted for swelling-prone clay formations.

The variation trends in the FEM and ML models for all parameters are in close agreement, with the ML model successfully capturing the shape, magnitude, and extent of the heave at the ground surface over the entire sensitivity range. 
Furthermore, the performance of the ML model demonstrates its ability to generalize over a wide range of parameter values without introducing significant biases or deviations, reinforcing its suitability for practical applications in site-specific analysis. 
The close fit also confirms the potential of the hybrid framework to accurately simulate complex nonlinear processes with reduced computational requirements.

\subsection{Uncertainty quantification}
\begin{figure}[h]
\centering
\includegraphics[width=0.4\textwidth]{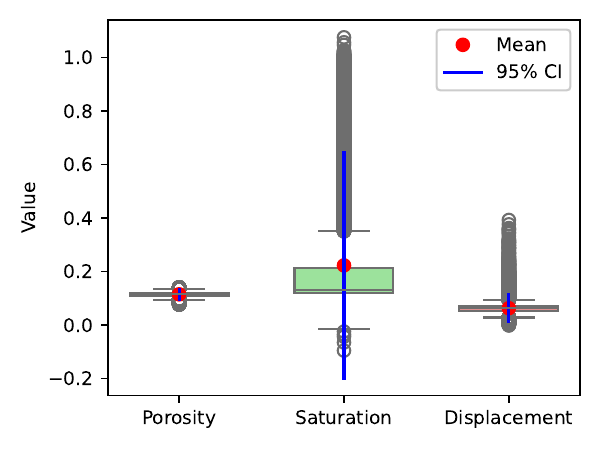}
\caption{Box plot of porosity, saturation and displacement with whiskers representing data spread, mean values marked in red, 95\% confidence intervals as blue lines, and outliers as individual points beyond the whiskers, indicating extreme variations in the data.}
\label{fig:box_plot_with_ci}
\end{figure}

The uncertainty associated with the porosity, saturation, and displacement predictions was evaluated using Monte Carlo simulations with a sample size of $N = \num{30000}$. 
The statistical metrics and corresponding confidence intervals derived from the analysis are summarized in Table~\ref{tab:uq_metrics} and Figure~\ref{fig:box_plot_with_ci}.

\begin{table}[h]
\caption{Comparison of uncertainty quantification metrics for marginalized predictions.}
\label{tab:uq_metrics}
\centering
\begin{tabular}{@{}lcccccccc@{}}
\toprule
 & Mean & Std. Dev. & COV & $q_1$ & $q_{25}$ & $q_{50}$ (Median) & $q_{75}$ & $q_{99}$ \\ 
\midrule
Porosity (-)   & $0.115$  & $0.009$  & $0.082$  & $0.086$  & $0.110$  & $0.116$  & $0.121$  & $0.137$  \\
Saturation (-)     & $0.223$  & $0.216$  & $0.968$  & $0.028$  & $0.120$  & $0.129$  & $0.212$  & $0.991$  \\
Displacement (\si{\meter}) & $0.063$  & $0.025$  & $0.399$  & $0.015$  & $0.053$  & $0.064$  & $0.069$  & $0.164$  \\ 
\bottomrule
\end{tabular}
\end{table}

Porosity has a mean of $0.115$ with a standard deviation of $0.009$ and a variance of approximately $0.00008$.
This relatively small variability is further reflected in the moderately narrow interquartile (IQR) range, where $q_{25} = 0.110$ and $q_{75} = 0.121$ are closely spaced around the median $q_{50} = 0.116$.
These results suggest that saturation predictions are stable, potentially due to the less dynamic nature of this rock property in the swelling layer, 
where spatial variations are relatively minimal. 
The upper outliers appear mostly beyond $q_{99} = 0.137$.

Saturation, by contrast, exhibits the highest variability among the three parameters, with a mean of $0.223$, a standard deviation of $0.216$, and a variance of $0.04666$.
This extensive spread is further evidenced by the wide IQR, where $q_{25} = 0.120$ and $q_{75} = 0.212$ are far away from the median $q_{50} = 0.129$.
These results indicate a high sensitivity of the saturation predictions to uncertainties in the input parameters,
reflecting its dynamic behavior.

Displacement has a mean of $0.063$, a standard deviation of $0.025$, and a variance of $0.00063$.
Although it displays greater variability than porosity, it remains more constrained than saturation.
The quantiles $q_{1} = 0.015$ and $q_{99} = 0.164$ indicate the influence of localized swelling effects and nonlinear deformations,
but the interquartile values $q_{25} = 0.053$ and $q_{75} = 0.069$ are relatively close to the median $q_{50} = 0.064$.

\section{Conclusion}

This study presents a hybrid framework based on a hydro-mechanical model previously developed to analyze the swelling of clay-sulfate rocks at the Staufen study site \citep{taherdangkoo2022coupled}. 
The framework predicts water front propagation and its effect on porosity and displacement with high accuracy, providing a balance between computational efficiency and prediction accuracy. 
The small residuals and lack of systematic bias demonstrate the model's reliability for the regression task. 
Sensitivity analysis shows the importance of site-specific parameters such as swelling pressure, permeability, and Young's modulus, emphasizing the importance of accurate site-specific characterization of material properties. 
Uncertainty quantification through Monte Carlo simulations confirms the robustness of the model and validates its suitability for long-term, nonlinear swelling processes.

A key advantage of this hybrid approach is the balance between computational efficiency and predictive accuracy. 
By using a machine learning surrogate, the framework enables fast simulations suitable for real-time applications and extensive parametric studies. 
The inclusion of penalties ensures that the predictions remain meaningful, making the framework adaptable to different geotechnical scenarios. 
The framework thus provides a valuable alternative for risk assessment, engineering design, and mitigation planning in regions affected by swelling. 
Overall, the work demonstrates the utility of machine learning as a computationally efficient approximation method in geotechnical applications. 

However, the reliance on synthetic data sets generated by FEM simulations may not fully capture real-world complexities, such as field-specific conditions. 
Extending the framework with physics-informed neural networks (PINNs) \citep{raissi2019physics}
to improve enforcement of physical constraints and continual learning could further enhance its capability for large-scale studies.
Future efforts could extend the framework to broader applications, such as tunneling, slope stability, and foundation design in geologically active areas.

\subsection*{Funding}
This research is funded by the German Research Foundation (DFG) for project TA 2076/1-1. 

\subsection*{Competing Interests}
The authors declare no competing interests.

\subsection*{Code Availability}
The machine learning code used in this study was developed using Jupyter Notebook and is available upon request from the corresponding author.

\subsection*{CRediT authorship contribution statement}
\textbf{Reza Taherdangkoo}: Conceptualization, Methodology, Validation, Formal analysis, Funding acquisition, Writing - Original Draft.
\textbf{Mostafa Mollaali}: Methodology, Formal analysis, Writing - Review \& Editing.
\textbf{Matthias Ehrhardt}: Methodology, Writing - Review \& Editing.
\textbf{Thomas Nagel}: Writing - Review \& Editing.
\textbf{Lyesse Laloui}: Writing - Review \& Editing.
\textbf{Alessio Ferrari}: Writing - Review \& Editing.
\textbf{Christoph Butscher}: Writing - Review \& Editing. All authors contributed to discussions of the results and their implications.

\bibliographystyle{apalike}

\end{document}